\documentclass[12pt]{article}
\def\er{{\cal R}}
\def\em{{\cal M}}
\def\l{\lambda}
\def\mn{{\mu\nu}}
\def\na{\nabla}
\def\I{{\rm I}}
\def\e{{\rm \, e}}
\def\sh{{\rm \, sinh}}
\def\ch{{\rm \, cosh}}
\def\th{{\rm \, tanh}}
\def\mo{{-1}}

\def\be{\begin{equation}}
\def\ee{\end{equation}}
\def\lb{\label}
\def\bea{\begin{eqnarray}}
\def\eea{\end{eqnarray}}

\def\PL#1{Phys.\ Lett.\ {\bf#1}}
\def\PRL#1{Phys.\ Rev.\ Lett.\ {\bf#1}}
\def\PR#1{Phys.\ Rev.\ {\bf#1}}
\def\NP#1{Nucl.\ Phys.\ {\bf#1}}
\def\JMP#1{J.\ Math.\ Phys.\ {\bf#1}}

\def\grq#1{{\tt gr-\allowbreak qc/\allowbreak#1}}
\def\hep#1{{\tt hep-\allowbreak th/\allowbreak#1}}

\begin{document}
\begin{titlepage}
\begin{flushright}
INFNCA-TH0203\\
\end{flushright}
\vspace{.3cm}
\begin{center}
\renewcommand{\thefootnote}{\fnsymbol{footnote}}
{\Large \bf The Cardy-Verlinde formula for 2D gravity}
\vfill
{\large \bf {M.~Cadoni$^{1,a}$\footnote{email: mariano.cadoni@ca.infn.it},
P.~Carta$^{1,a}$\footnote{email: paolo.carta@ca.infn.it}
and S.~Mignemi$^{2,a}$\footnote{email: salvatore.mignemi@ca.infn.it}}}\\
\renewcommand{\thefootnote}{\arabic{footnote}}
\setcounter{footnote}{0}
\vfill
{\small
$^1$ Universit\`a di Cagliari, Dipartimento di Fisica,\\
Cittadella Universitaria, 09042 Monserrato, Italy\\
\vspace*{0.4cm}
$^2$ Universit\`a di Cagliari, Dipartimento di Matematica,\\
Viale Merello 92, 09123 Cagliari, Italy\\
\vspace*{0.4cm}
$^a$ INFN, Sezione di Cagliari}\\
\vspace*{0.4cm}
\end{center}
\vfill
\centerline{\bf Abstract}
\vfill

We discuss the different bounds on entropy in the context of
two-di\- mensional cosmology. We show that in order to obtain
well definite bounds one has to use  the scale
symmetry  of the gravitational theory. We then extend the recently found
relation between the Friedmann equation and the Cardy formula
to the case of two dimensions. In particular, we find that in two
dimensions this relation requires that the central
charge $c$ of the conformal field theory
is given in terms of the Newton constant $G$ of the
gravitational theory by $c=6/G$.
\vfill
\end{titlepage}
%


\section{Introduction}

In a recent paper \cite{ver}, Verlinde has discussed the cosmological
bounds on entropy for spacetimes of dimension $d>2$.
These are based on the holographic principle \cite{tH}, which states
that the entropy contained into a given region of space should be
bounded by the area of the spacelike surface that encloses it.
Another important result of \cite{ver} is that the Friedmann
equation of cosmology for a radiation-dominated universe can be shown
to be equivalent to the Cardy formula for the entropy of a conformal
field theory describing the radiation. This observation has of course
important implications, which have not been fully clarified yet.

The context of the original proposal of  Verlinde was
$d>2$ cosmology. Although the Cardy-Verlinde formula has been generalized to
describe other gravitational systems \cite{gruppo1,gruppo2}, in particular
black holes,
a discussion of the  $d=2$ case is still lacking.
The two-dimensional (2D) limit of the Cardy-Verlinde proposal is
interesting for various reasons.  From  investigations of the
anti-de Sitter (AdS)/Conformal Field Theory (CFT) correspondence,
we know that there are 2D
gravitational systems that admit 2D CFTs
as duals \cite{CM99,cadcav}. In this case one can make direct use of the
original Cardy formula \cite{cardy} to compute the entropy
\cite{CM99,cadcav}.
A comparison of these results with a 2D generalization of the
Cardy-Verlinde formula could be very useful in particular for the
understanding of the puzzling features of the AdS/CFT correspondence in
two dimensions \cite{strominger}.
Another point of interest in extending the Cardy-Verlinde formula to
$d=2$ is the clarification of the meaning of the holographic principle
for 2D spacetimes.
The boundaries of spacelike regions of 2D spacetimes are points, so that
even the notion  of holographic bound is far from trivial.

A generalization of the work of Verlinde to two spacetime dimensions
presents several difficulties, essentially for dimensional reasons.
First of all, in two dimensions one cannot establish
a area law, since black hole horizons are isolated points.
Moreover, the spatial coordinate is not a "radial" coordinate and
hence one cannot impose a natural normalization on it. As we shall
see later on this paper in detail, this fact is
connected, at least for the 2D gravity model we consider here,
to a  scale symmetry of 2D gravity \cite{CC}. Related
to this symmetry is also the fact that the 2D
gravitational coupling constant $G$ is dimensionless,
and hence one cannot even define a "Planck" length.
Finally, if one works, as we do in this paper, in the context
of scalar-tensor theories of gravity,  the  2D cosmological equations
are quite different from their  Friedmann-Robertson-Walker $d>2$
counterparts.

Some of these problems may of course be solved if one considers
gravity in two dimensions as a dimensionally reduced theory.
However, if one wants to keep a purely two-dimensional point of view,
one has to deal with the particular features of 2D gravity.

In this paper, we wish to extend Verlinde's results to a
radiation-dominat\-ed 1+1 universe, in which the gravitational interaction
is governed by a Jackiw-Teitelboim (JT) model \cite{JT,cosm}.
We shall see that this goal can be achieved,
provided that some free parameters appearing in the solutions
are fixed using the scale symmetry of the theory.

The paper is organized as follows: in sect. 2 we discuss the
cosmological model derived from the JT model. In sect. 3 we
discuss how the standard cosmological bounds on the entropy
can be generalized to our case. In sect. 4 we use the scale symmetry
of the gravitational theory to fix the free dimensionless parameters
appearing in the bounds.
In sect. 5 we investigate the relations between the cosmological
equations and the Cardy formula. The cosmological bounds on
the temperature of a radiation-dominated universe are discussed in sect.
6. Finally, in sect. 7, we present our conclusions.

\section{Two-dimensional cosmology}

Let us consider the action for  JT gravity minimally  coupled to
matter,
\be\lb{action}
I=\int d^2x\sqrt{-g}\left(\eta{\er-2\l^2\over 16\pi G}+L_M\right),
\ee
where $\eta$ is a scalar field (the dilaton), $G$ is the dimensionless
2D Newton constant, which could be absorbed in a redefinition of the
dilaton, $\l^{2}$ is a cosmological constant  and $L_{M}$ is the
matter lagrangian.
We want to discuss a radiation-dominated 1+1 universe,
in which case $L_{M}$ describes free (or weak interacting) massless
particles. In general, $L_{M}$  can be given in terms of a 2D  CFT.
Because the matter lagrangian is that of a perfect fluid, it can be
taken proportional to its density, $L_M=-\rho$ \cite{DC}.
A constrained variation of (\ref{action}) gives the field
equations \cite{cosm}
\bea\lb{feqa}
&&\er=2\l^2,\nonumber\\
&&-(\na_\mu\na_\nu-g_\mn\na^2)\eta+\l^2g_\mn\eta=8\pi T_\mn,
\eea
where $T_\mn$ is the standard energy-momentum tensor of a
perfect fluid, $T_\mn=p_Mg_\mn+(\rho+p_M)u_\mu u_\nu$, with $p_M$
the pressure of the fluid.
The field equations (\ref{feqa}) tell us that, independently of the
matter, the spacetime has constant, positive  curvature.
It is therefore given by a 2D de Sitter (dS) spacetime.

We make the ansatz $ds^2=-dt^2+R^2(t)dx^2$, $\eta=\eta(t)$, with
$0\le x\le2\pi$. Since we take $x$ periodic, we are considering a
closed 1+1 universe. However, our considerations can be easily
extended to a open universe.
The field equations then take the form
\bea
&&\ddot R-\l^2R=0\lb{feq},\nonumber\\
&&\dot R\dot\eta-\l^2R\eta=8\pi GR\rho,\nonumber\\
&&\ddot\eta-\l^2\eta=-8\pi Gp_M.
\eea

Combining the field equations, one obtains the energy momentum
conservation in the form $\dot\rho=-(p_M+\rho)\dot R/R$.
For a perfect fluid, $p_M=\gamma\rho$, and this relation can be
integrated to yield $\rho R^{1+\gamma}=$ const $= M/2\pi$.
We are considering the case in which the matter is constituted
of pure radiation, for which $\gamma=1$, so that we have,
\be\lb{eqs}
\rho={M\over 2\pi R^2}.
\ee

The general solution of the first of Eqs. (\ref{feq}) is
\be
R(t)=\bar a\e^{\l t}+\bar b\e^{-\l t}.
\ee
Depending on the relative sign of the integration constants
$\bar a$ and $\bar b$, and with a suitable choice of the origin
of time, the solution can assume three qualitatively different forms:

\bea
&\I)&R={a\over\l}\e^{\l t},\lb{met1}\\
&\I\I)&R={a\over\l}\sh\l t,\\
&\I\I\I)&R={a\over\l}\ch\l t.\lb{met3}
\eea
where $a$ is a dimensionless parameter
(we choose this normalization, in order to give to $R$ the physical
dimension of a length).

The solutions for the scalar field are, respectively \cite{cosm}
\bea\lb{dilaton}
&\I)&\eta=\eta_0\e^{\l t}-{4GM\over 3a^2}\e^{-2\l t},\nonumber\\
&\I\I)&\eta=\eta_0\ch\l t+{4GM\over a^2}\left(1+\ch\l t\ \log\th{\l t\over2}
\right),\nonumber\\
&\I\I\I)&\eta=\eta_0\sh\l t-{4GM\over a^2}\left(1+
\sh\l t\ \arctan\sh\l t\right),
\eea
with $\eta_0$ an integration constant. All solutions are of course
locally isomorphic to de Sitter spacetime, but with different
parametrization, covering different regions of the 2D manifold.
In particular, I and II possess a horizon at
$t=-\infty$ and $t=0,$ respectively.
Moreover for all solutions I, II and III the scalar field  $\eta$ has a
zero at a finite value $t_{0}$ of the cosmological time $t$.
Since we are dealing with a Brans-Dicke-like theory of gravity,
$\eta^{-1}$ represents a time-dependent effective Newton constant.
The instant $t=t_{0}$ may therefore be interpreted as an initial
singularity and we will restrict ourselves to consider only times
$t\ge t_{0}$, when $\eta\ge 0$.

In two dimensions there is no direct analog of the $d$-dimensional
Friedmann equation
\be\lb{fried}
H^2=\l^2+{16\pi G\over(d-1)(d-2)}\rho-{1\over R^2},
\ee
where $H\equiv\dot R/R$ is the Hubble parameter (Notice that the
Friedmann equation (\ref{fried}) is singular for $d=2$).
However, an equation
for $H^2$ can be obtained by integrating the first equation in (\ref{feq}):

\be\lb{hubble1}
H^2=\l^2-{\alpha\over R^2}.
\ee
The constant of integration $\alpha$ for the solutions I-III
is respectively, 0, $-a^2$, $a^2$.
For a radiation-dominated universe, one can use Eq. (\ref{eqs}) into
Eq. (\ref{hubble1}),
to obtain an expression
formally similar to (\ref{fried}),
\be\lb{hubble}
H^2=\l^2+8\pi G\rho-{\bar\alpha\over R^2},
\ee
where $\bar\alpha=\alpha+4GM$. Notice that we are using the
arbitrariness of the integration constant $\alpha$  to make the metric
of the spacetime dependent on the matter. Consistently with the
field equations  (\ref{feq}) the effect of the matter on the metric
can be only encoded in the choice of  integration constants.

Our gravity model (\ref{action}) has a cosmological constant
different from zero.  Therefore, one can assign to the vacuum an energy
$E_{\lambda}= {\lambda^{2}R/ 4G}$ and a pressure $p_\l=-E_\l$.
This permits to write Eq. (\ref{hubble}) in terms of the total energy
$E=(E_{\lambda}+E_{M})$, where $E_{M}=M/R$ is the energy of the matter,
\be\lb{hubble1a}
H^2= {4G E\over R}-{\bar\alpha\over R^2}.
\ee

\section{Entropy bounds}
In $d>2$ a bound on the entropy of a macroscopic system, $S\le S_B$, is
believed to hold, where the Bekenstein entropy $S_B$ is defined as \cite{bek}
\be
S_B={2\pi\over d-1}ER,
\ee
with $E$ the total energy and $R$ the linear size of the
system. This bound is verified for standard gases, but the
numerical factor in front of $ER$ is fixed by the assumption that
the bound is saturated by black holes.

A generalization of this bound to two dimensions is not
straightforward.
Consider for example the 2D anti-de Sitter black hole
\cite{CM}, which  is a solution of the gravity model (\ref{action})
with  negative $\l^{2}$,
\be\label{bh}
ds^2=-(\l^2x^2-m^2)dt^2+(\l^2x^2-m^2)^\mo dx^2\qquad \eta=\eta_0\l x
\ee

A horizon occurs at $x_0= m/\l$, and one can associate to it the
temperature $T=\l m/2\pi$ and the entropy $S=2\pi\eta_0m$. Moreover, by
standard methods, one can assign to the black hole the ADM mass
$\em=\eta_0\l m^2/2$. Thus one gets the relation
\be
S={4\pi\em x_0\over m^2}
\ee

If one identifies the energy $E$ of the black hole with its ADM mass
and its size $R$ with the length $x_0$, one finds $S\propto ER$,
but the ratio $S/ER$ grows without limit for small $m$
(in the limit $m\to0$, however,
all quantities vanish).
The same situation occurs in more general two-dimensional models.
The problem is connected to the fact that in two dimensions there
is no radial coordinate and hence the coordinate $x$ cannot be
properly normalized, or equivalently to the scale symmetry of
the model, which will be discussed in the next section.

Thus, although one can envisage a Bekenstein bound of the form
\be\lb{beke}
S\le S_B=2 \pi\epsilon ER,
\ee
which can also be deduced from the
thermodynamics of a one-dimensional gas,
the coefficient $\epsilon$ is not clearly determined.
Notice that in the radiation-dominated cosmological model of
the previous section
$S_B=2\pi M$ is a conserved quantity, proportional to the matter density.

The Bekenstein bound is believed to hold when the gravitational energy of
the system is small with respect to its total energy, i.e in a weak-gravitating
regime. For strong-gravitating
systems, i.e. systems for which $HR>1$, a different bound must be introduced.
In order to establish in which regime a given system is, it is useful
to define a Bekenstein-Hawking entropy $S_{BH}$ as the Bekenstein
entropy of a system with $HR=1$ \cite{ver}.
From the "Friedmann" equation (\ref{hubble1a}) one obtains
\be\lb{e}
S_{BH}={V\over4GR}(1+\bar\alpha)={\pi\over2G}(1+\bar\alpha),
\ee
where $V=2\pi R$ is the spatial volume.
Recalling the definition of $\bar\alpha$, Eq. (\ref{e}) can also be written
as
\be\lb{sbh}
S_{BH}={\pi\over2G}(1+\alpha)+2\pi M
\ee
This is the sum of two constant contributions: the first depends only
on the geometry, while the second, which is proportional to $S_B$, depends
only on the matter content and is not present in $d>2$.
In absence of matter, the second contribution vanishes.
The appearance of a factor $\alpha$ proportional to $a^2$ is again
a consequence of the scale invariance of the theory, which does
not fix the scale of the spatial coordinate in the solutions.

Notice that the bound $S\le S_{BH}$ is a truly holographic bound for
a 2D spacetime. The boundary of a spatial section of a 2D
universe are two points: thus the holographic principle states that the
entropy can only depend on the Newton constant $G$ and on a
dimensionless parameter.

In a cosmological contest and  when $HR>1$ the Bekenstein bound must be replaced
by a holographic bound. However, it has been argued that the
Bekenstein-Hawking bound $S\le S_{BH}$ is not the right choice.
A suitable bound is given by
the Hubble entropy $S_H$, defined as the entropy of a
universe filled with black holes of the size of a particle horizon
\cite{FS}. Later, a weaker definition of $S_H$ was proposed, in which the
maximal size of the black holes is the Hubble  radius $H^\mo$ \cite{EL}.
In our 2D context, $S_H$ can be calculated as follows \cite{ven}.
From Eq. (\ref{e}), a black hole of radius $H^\mo$ has entropy
$(1+\bar \alpha)HV_H/4G$. Since the universe can contain $N_H=V/V_H$
black holes, one obtains
\be\lb{shubble}
S_{H}={VH\over4G}(1+\bar \alpha)={\pi RH\over 2G}(1+\bar \alpha)
\ee
Notice that, while the solution I and II have a particle horizon
of the size of the Hubble radius, in case III the size of the
particle horizon grows exponentially with time.

To conclude,  although in the 2D case one may be able to obtain
different entropy bounds, they do not seem to be universal.
The entropy bounds $S\le S_{BH}$ and
$S\le S_{H}$ depend
on the arbitrary, dimensionless parameters $\bar\alpha$ and $\epsilon$.
They  appear to be defined up to arbitrary
scales. In the next section we will show that this fact is a
consequence of a scale symmetry of a our 2D gravity model.
This scale symmetry is a peculiarity of two-dimensional gravity and
is related to the impossibility of defining an area law for the entropy.

\section{Scale symmetry and entropy bounds}

It is well known that 2D  AdS space has $SL(2,R)$ as
isometry group (see for instance Ref. \cite{CM99}).  The spacetime
metric is
therefore invariant under the subgroup of $SL(2,R)$ describing
dilatations, which for the ground state $m=0$ in Eq. (\ref{bh}) is
realized as
\be\lb{dil}
x\to\nu x,\quad t\to t/\nu.
\ee
Under this scale transformation
the dilaton $\eta$ is not invariant, $\eta \to \nu \eta$, but the scale
factor $\nu $ can be absorbed in a different definition of the
integration constant $\eta_{0}$ appearing in Eq. (\ref{bh}). It is
evident that this scale transformation is a classical  symmetry of
the theory because under $\eta \to \nu \eta$ the action for pure
gravity changes just for an overall constant factor.

This scale symmetry is also a classical invariance of our (de Sitter)
action (\ref{action})  in absence of matter.
In the matter-coupled case the scale transformation just changes by
a constant factor the Newton constant $G$.

It is not difficult to realize that the presence of the integration
constants $\eta_{0}, m$
(in the AdS  solution)   and of $\eta_{0}, a$ (in the dS solution) is
a consequence of  the scale symmetry. The transformation
(\ref{dil}) maps one solution of the fields equations characterized by
$\eta_{0}, m$ into an other solution with  different
values of the integration constants, $\eta_{0}', m'$.
It is therefore evident that we can use the scale symmetry (\ref{dil})
to write the  entropy bounds in a form that is independent of the
dimensionless parameters $\bar\alpha, \epsilon$.

Instead of directly working on the cosmological solution,
it is more instructive to fix these parameters by considering
the two-dimensional black hole.
Since the 2D cosmological solutions are the analytical continuation
$\lambda \to i\lambda$ of the black hole solutions (\ref{bh}), one can
 fix $\bar\alpha, \epsilon$ using the latter as the maximum  entropy
configuration.

The problem reduces then to fix the dependence on $m$ of the
thermodynamical parameters $E={\cal M}, T, S$ of the AdS$_{2}$
black hole (\ref{bh}). 
Introducing the length scale
$L=1/\lambda$  and the central charge $c=
12\eta_{0}$ of the thermal CFT  arising in the AdS$_{2}$/CFT
correspondence one has \cite{CM99,CC}
\be\lb{TSE}
T={1\over 2\pi} {m\over L},\quad S=2\pi\,{c\over 12} m\quad E= {c\over
24}
{m^{2}\over  L}.
\ee
For a generic black hole solution the behavior under the scale
transformations has been given in Ref \cite{CC}.
The AdS$_{2}$ black hole metric (\ref{bh}) is invariant under the
scale transformations,
\be\lb{scat}
x\to \nu x,\quad t\to {t\over \nu}, \quad {\cal  M}\to\nu^{2}{\cal M}.
\ee
The dilaton transforms as
\be\lb{dila}
\eta\to \nu \eta,
\ee
whereas, $T,S,E$ scale as
\be\lb{scale}
T\to \nu T,\quad S\to\nu S,\quad E\to \nu^{2}E.
\ee

The physical meaning of this scale invariance of the theory
can be easily understood.
It is a general feature of all  metric theories of gravity that
lengths (or masses) can be only measured with reference to an
(asymptotic) reference frame.
For asymptotically Minkowskian solutions this frame is given by
Minkowski space with the usual normalization ($ds^{2}= -dt^{2}+dx^{2}$).
Owing to the dilatation isometry of AdS$_{2}$, for solutions which are
asymptotically AdS there is no such ``preferred'' reference frame.
We are free to change the length of our rule using the scale
transformations (\ref{scat}), without changing the physics.
We have a sort of gauge symmetry, stating that black hole solutions
connected by the scale transformations (\ref{scat}) are physically
equivalent\footnote{Also the geodesic motion is unaffected by a
rescaling of $m$, which simply shifts the definition of the energy of
a test particle.}.  However, the energy $E$ and entropy $S$ change under the scale
transformation, they are not gauge-invariant quantities.
For this reason, although  we cannot  find an absolute  upper bound for $S$,
every $m$-dependent  bound of the form $S(m)\le S_{H}(m)$ has a
gauge-invariant meaning.
Thus, fixing the gauge we can remove from the entropy bounds
the dependence on the dimensionless parameters $\alpha, \epsilon, m$.
This  can be easily done by using  Eqs. (\ref{scale}) with $\nu=1/m$ into Eqs.
(\ref{TSE}) to remove the $m$-dependence of $T,S,E$,
 \be\lb{TSE1}
T={1\over 2\pi} {1\over L},\quad S={c\over 12} 2\pi \quad E= {c\over
24}
{1\over  L},
\ee
and choosing the value of $m$ to fix, by means of Eq. (\ref{dila}),
$\eta_{0}$ (hence the
central charge $c$) in terms of the Newton constant $G$,
\be
\eta_{0}= {c\over 12}= {1\over 2G}.
\ee
With this choice, one has for the two-dimensional black hole
\be
S_B=2\pi ER,
\ee
fixing $\epsilon=1$ in the Bekenstein bound (\ref{beke}).
Moreover, if one identifies $S_{BH}$  with the entropy of the AdS$_{2}$ black hole
given by Eq. (\ref{TSE1}), one
finds $\bar \alpha=1$. It follows immediately
\be\lb{BH}
S_{BH}= {\pi\over G},
\ee
and hence
\be\lb{H}
S_{H}={\pi R H\over G}.
\ee
With these definitions, one has the relation
\be
S_H^2=S_{BH}(2S_B-S_{BH}).
\ee
Once the value of  $\bar \alpha$ has been fixed,  $\bar \alpha=1$
the cosmological equation (\ref{hubble1a}) takes the Friedmann form
\be\lb{hubble2}
H^2= {4G E\over R}-{1\over R^2}.
\ee

\section {Cardy-Verlinde formula for 2D cosmology}

Recalling the definition of the Bekenstein-Hawking entropy,
one can define a Bekenstein-Hawking energy $E_{BH}$, as the energy
corresponding to the condition $S_B=S_{BH}$, at which the gravitational
system becomes strong-coupled:
\be\lb{EBH}
E_{BH}={1\over 2RG}
\ee
Using Eqs. (\ref{H}),(\ref{EBH}) and (\ref{hubble2}) one gets a first
form of the 2D Cardy-Verlinde formula
\be\lb{verlinde}
S_{H}= 2\pi R \sqrt{E_{BH}(2E-E_{BH})}
\ee

Following Verlinde \cite{ver} it is now useful to split the total energy $E$
in extensive $E_{E}$ and non-extensive (Casimir)  part $E_{C}/2$
\footnote{The Casimir energy
is of course $E_C/2$. We adopt this rather odd notation in order to
conform to ref. \cite{ver} and to simplify some of the following formulas.}
\be\lb{p1}
E=E_{E}+{E_{C}\over2},
\ee
where $E_C$ is defined as
\be\lb{p3}
E_{C}\equiv E+pV -TS
\ee
with $p=p_M+p_\l$.

The scaling behavior of $E_{C}$ follows from general arguments
(for instance the form of the Casimir energy of a CFT on the
cylinder),
$E_{C}(\Lambda S, \Lambda V)=\Lambda^{-1}¥ E_{C}( S, V)$.
Moreover, conformal invariance implies that $ER$ is independent of the volume
$V=2\pi R$. Combined with the scaling behavior of $E_{E}$, which by
definition is $E_E(\Lambda S, \Lambda V)=\Lambda E_{E}( S, V)$,
this gives
\be
E_{E}= {b\over 4\pi R} S^{2}, \quad E_{C}= {d\over 2\pi R}
\ee
where $b,d$ are arbitrary constants. Using this equation and
(\ref{p1}) one gets
\be\lb {p3a}
S= {2\pi R\over \sqrt{bd}} \sqrt{E_C(2E-E_{C})}
\ee

Eq. (\ref{p3a}) becomes the Cardy formula if we
take $bd=1$ and consider  a 2D CFT on the cylinder. In fact,
in this case the total and Casimir energy are given by
\be
E={l_{0}\over R},\quad {E_{C}\over 2}= {c\over 24 R},
\ee
where $c$ and $l_{0}$ are the central charge and the eigenvalue of the
Virasoro operator $L_{0}$ of the CFT and $R$ is the radius of
the cylinder.
Inserting the previous equations into Eq. (\ref{p3a}) one finds the Cardy
formula
\be\lb{cardy}
S=2\pi\sqrt{{c\over 6}\left(l_{0}- {c\over 24}\right)}.
\ee

Comparing Eq. (\ref{verlinde}) with  Eq. (\ref{p3a}) ($bd=1$)
we find that they agree if we take $S=S_{H}$ and $E_{BH}=E_{C}$.
We are therefore led to a cosmological bound for the Casimir
energy
\be\lb{newbound}
E_{C}\le E_{BH},
\ee
analogous to the one proposed for higher-dimensional cosmology \cite{ver}.
Using the Friedmann Equation (\ref {hubble2}), one easily finds that for
$HR\ge 1$, $E\ge E_{BH}$. Hence for strongly gravitating systems we have
\be
E_{C}\le E_{BH}\le E.
\ee
This bound shares some nice features with its higher-dimensional counterpart:
(a) It is always valid and does not break down for $HR\le 1$.
(b) Its physical meaning is that the cosmological energy by itself is not
sufficient to produce a black hole of the size of the entire universe.
In fact, the AdS$_{2}$ black hole saturates the bound.
(c) For $HR>1$ it is equivalent to the Hubble bound $S<S_{H}$.
(d) When the bound is saturated $E_C=E_{BH}$ and the cosmological
equation (\ref{hubble2}) becomes the Cardy formula (\ref{cardy}).
The translation table  between 2D cosmology  and 2D CFT is given
by
\bea
{c\over 12}&\longleftrightarrow &\,{1\over 2G},\nonumber\\
l_{0}&\longleftrightarrow&\, ER,\nonumber\\
S&\longleftrightarrow&\, {\pi RH\over G}.
\eea

\section {Limiting temperature}

The cosmological equation  (\ref{feq}) can be used, in conjunction with
Eq. (\ref{hubble2}), to give a lower bound
for temperature in a radiation-dominated universe. From (\ref{feq}) and
(\ref{hubble2}) follows
\be\lb{hubbledot}
\dot H = \lambda^{2} -{4GE\over R} +{1\over R^{2}}={1-4GM\over R^2}.
\ee
Defining the Hubble temperature
\be
T_{H}=-{1\over 2\pi} {\dot H\over H},
\ee
and using the definition of $S_{H}$ and $E_{BH}$,
Eq. (\ref{hubbledot}) becomes
\be\lb{p6}
E_{BH}= -T_{H}S_{H}+ E+pV.
\ee

Comparing (\ref{p6}) with (\ref{p3}) and using the bounds $S\le S_{H}$,
$E_{C}\le E_{BH}$ one obtains a lower bound for $T$,
\be
T\ge T_{H}.
\ee

\section{Conclusions}
We have shown that the analysis of Verlinde on  cosmological entropy
bounds and their relations with CFT can be extended to a
two-dimensional model, in spite of the difficulties related to the
definition of a holographic principle in two dimensions.
The identification of the Friedmann equation with the Cardy formula requires
the use of the scaling invariance of the theory in order to fix
some dimensionless parameters.

The most striking feature of the mapping between 2D cosmology and 2D
CFT is the identification of the Newton constant in terms of the central charge
of the CFT. The correspondence between cosmological equations and the
Cardy formula requires  $c=6/G$. This relation has an obvious
holographic nature. In higher dimensions the holographic principle
requires $c\propto V/GR$. Extended to the 2D case  where $V=2\pi R$
this relation reproduces our result.
Further support to the holographic origin of this relation, comes
from the fact that it can also be deduced using the  AdS/CFT
correspondence in two dimensions \cite{CM99,cadcav,CCKM}.
In this context, it has been found that the  central charge
of the CFT living on the boundary
of AdS$_{2}$ is given by $c=12\eta_{0}$, which is our present result,
because $\eta_{0}$, is proportional to  the inverse of the 2D Newton constant.
Owing to the dilatation symmetry of our model, under which $\eta_{0}$
scales as in Eq. (\ref{dila}), the coefficient of proportionality
between $\eta_{0}$  and $G^{-1}$ depends on the dilatation-gauge we choose.
\eject

\end{document}